\documentstyle[12pt]{article}

\begin{document}
\hspace{80mm}  Mod. Phys. Lett. A 13(1998)33.

\begin{center}
\bf{  Quantum fluctuations of the angular momentum and
                energy of  \\ the ground state  }        \\
\vspace{3mm}
\rm                   {   M.N. Sergeenko   }                \\
\vspace{2mm}
\it{ The National Academy of Sciences of Belarus, Institute of Physics,\\
                      Minsk 220072, Belarus \ and           \\
             Gomel State University, Gomel 246699, Belarus }
\end{center}

\begin{abstract}
Quasiclassical solution of the three-dimensional Schr\"odinger's
equation is given. The existence of nonzero minimal angular 
momentum $M_0 = \frac\hbar 2$ is shown, which corresponds to the 
quantum fluctuations of the angular momentum and contributes to
the energy of the ground state.
\\  ~
\\
\noindent PACS number(s): 03.05.Ge, 03.65.Sq   \end{abstract}

One of the fundamental features of quantum mechanical systems is
nonzero minimal energy which corresponds to zero oscillations. The
corresponding wave function has no zeros in physical region. Typical
example is the harmonic oscillator.

The eigenvalues of the one-dimensional harmonic oscillator are 
$E_n =\hbar\omega_0(n+\frac 12)$, i.e. the energy of zero oscillations
$E_0 =\frac 12 \hbar\omega_0$. In three-dimensional case, in the
Cartesian coordinates, the eigenvalues of the oscillator are $E_n =
\hbar\omega_0 (n_x+n_y+n_z+\frac 32)$ \cite{Flu}, i.e. each degree
of freedom contributes to the energy of the ground state, $E_0 =
E_{0,x} + E_{0,y} + E_{0,z} = $ $\frac 32\hbar\omega_0$.

Energy of the ground state should not depend on coordinate system.
This means that, in the spherical coordinates, each degree of freedom
(radial and angular) should contribute to the energy of zero
oscillations. In many applications and physical models a nonzero
minimal angular momentum $M_0$ is introduced (phenomenologically) in
order to obtain physically meaningful result \cite{Iwa}. However, the
existence of $M_0$ follows from the quasiclassical solution of the
three-dimensional Schr\"odinger's equation.

Known exact methods to solve the Schr\"odinger's equation are usually
mathematical methods. However, together with quantum mechanics,
the appropriate method to solve the Schr\"odinger equation has been
developed; it is general for all types of problems in quantum
mechanics, and its correct application results in the exact
energy eigenvalues for {\em all} solvable potentials. This is the
phase-space integral method which is also known as the WKB method
\cite{Fro,Tr}.

The general form of the semiclassical description of
quantum-mechanical systems has been considered in Ref. \cite{Mi}.
It was shown that the semiclassical description resulting from
Weyl's association of operators to functions is identical with the
quantum description and no information need to be lost in going from
one to the another. What is more "the semiclassical description is
more general than quantum mechanical description..."  \cite{Mi}. The
semiclassical approach merely becomes a different representation of
the same algebra as that of the quantum mechanical system, and then
the expectation values, dispersions, and dynamics of both become
identical.

The WKB method was originally proposed for obtaining approximate
eigenvalues of one-dimensional and radial Schr\"odinger problems in
the limiting case of large quantum numbers. Exactness of the method
for the solvable potentials has been proved in many works
\cite{Fro,Ros}. As for multi-dimensional problems, the quasiclassical
method is even more efficient and predictive.

In the quasiclassical method, the classic quantities such as classic
momentum, classic action, phase, etc. are used. The WKB quantization
condition and WKB solution are writing via classic momentum.
However, the generalized moments obtained from the separation of the
three-dimensional Schr\"odinger equation are different from the
corresponding classic moments.

The standard WKB method in leading order in $\hbar$ {\em always}
reproduces the exact spectrum for the solvable spherically symmetric
potentials $V(\vert\vec r \vert)$ if the Langer correction
$l(l+1)\rightarrow (l+\frac 12)^2$ \cite{Lang} in the centrifugal
term of the radial Schr\"odinger equation has fulfilled. The ground
of this correction for the special case of the Coulomb potential was
given by Langer ($1937$) \cite{Lang} by means of reducing the
Schr\"odinger equation to canonical form (without first derivatives).

However, the Langer replacement $l(l+1)\rightarrow (l+\frac 12)^2$ is
universal for any spherically symmetric potential and requires
modification of the squared angular momentum. The Schr\"odinger's
equation for a spherically symmetric potential $V(r)$, in the
representation of the wave function $\psi(\vec r)$, assumes
separation of variables [$\psi(\vec r) = R(r)\Theta(\theta)
\Phi(\varphi)$]. After excluding the first derivatives, this equation 
can be written in the form of the classic equation

\begin{eqnarray}
-\hbar^2\frac{\tilde R_{rr}''}{\tilde R} +
\frac 1{r^2}\left(-\hbar^2\frac{\tilde\Theta_{\theta\theta}''}
{\tilde\Theta} - \frac{\hbar^2}4\right) +
\frac 1{r^2\sin^2\theta}\left(-\hbar^2\frac{\tilde
\Phi_{\varphi\varphi}''}{\tilde\Phi} -
\frac{\hbar^2}4\right) =     \\
 2m[E - V(r)],   \nonumber          \end{eqnarray}
where $\tilde R(r) = rR(r)$, $\tilde\Theta(\theta) =
\sqrt{\sin(\theta)}\Theta(\theta)$, $\tilde\Phi(\varphi) =
\Phi(\varphi)$. Separation of equation (1) results in the three
second-order differential equations in canonical form:

\begin{equation}
\left(-i\hbar\frac d{dr}\right)^2\tilde R =
\left[2m(E-V) - \frac{\vec M^2}{r^2}\right]\tilde R, \end{equation}

\begin{equation}
\left[\left(-i\hbar\frac d{d\theta}\right)^2 - \left(\frac\hbar
2\right)^2\right]\tilde\Theta(\theta) = \left(\vec M^2 -
\frac{M_z^2}{\sin^2\theta}\right)\tilde\Theta(\theta),
\end{equation}

\begin{equation}
\left[\left(-i\hbar\frac d{d\varphi}\right)^2 - \left(\frac\hbar
2\right)^2\right]\tilde\Phi(\varphi) = M_z^2\tilde\Phi(\varphi),
\end{equation}
where $\vec M^2$ and $\vec M_z^2$ are the constants of separation
and, at the same time, integrals of motion.  Equations (2)-(4)
have the quantum-mechanical form $\hat f\psi = f\psi$, where
$f$ is the physical quantity and $\hat f$ is the corresponding
operator. Generalized moments in right-hand sides of Eqs. (2)-(4)
should be used in the WKB quantization condition and WKB solution and
solve the known problem of application of the WKB method to the
three-dimensional Schr\"odinger's equation\footnote {Quantities
$-\hbar^2\tilde R_{rr}''/\tilde R$,
$-\hbar^2\tilde\Theta_{\theta\theta}''/\tilde\Theta$,
$-\hbar^2\tilde\Phi_{\varphi\varphi}''/\tilde\Phi$ obtained after
separation of equation (1) are usually considered as the squared
moments which are used in the WKB quantization condition, that results
in the known difficulties of the WKB method.}.

We solve each of them obtained after separation equation by the same 
method, i.e. the WKB method. For the projection of the angular momentum,
from the quantization condition $\oint p(\varphi)d\varphi =
2\pi m\hbar$ (here $p(\varphi) = M_z$), we have $M_z = m\hbar$, $m =
0,1,2,...$. The squared angular momentum eigenvalues, $\vec M^2$, are
defined from the WKB quantization condition ($\theta_1$ and $\theta_2$
are the classical turning points)

\begin{equation}
\int _{\theta_1}^{\theta_2}\sqrt{\vec M^2 - \frac{M_z^2}
{\sin^2\theta}}d\theta = \pi\hbar\left(n_\theta +\frac 12\right), \ \ \
n_\theta = 0,1,2,...     \end{equation}
Integration of (5) gives for $\vec M^2$,

\begin{equation}
\vec M^2 = \left(l +\frac 12\right)^2\hbar^2, \ \ \ l = |m| + n_\theta.
\end{equation}
Energy eigenvalues are defined from the condition

\begin{equation}
\int_{r_1}^{r_2}\sqrt{2m[E - V(r)] - \frac{\vec M^2}{r^2}}dr =
\pi\hbar\left(n_r +\frac 12\right), \ \ \  n_r = 0,1,2,...,   \end{equation}
where $r_1$, $r_2$ are the classical turning points.

The WKB solution corresponding to the eigenvalues (6) has the correct
asymptotic behavior at $\theta \rightarrow 0$ and $\pi$ for all
values of $l$. In the representation of the wave function $\psi (\vec
r)$, $\Theta_l^m(\theta) = \tilde\Theta^{WKB}(\theta)/
\sqrt{\sin\,\theta}\propto\theta^{|m|}$ which corresponds to the
behavior of the exact wave function $Y_{lm}(\theta ,\varphi )$ at
$\theta\rightarrow 0$. The normalized quasiclassical solution [far
from the turning points, where $p(\theta)\simeq (l+\frac 12) \hbar$]
in the representation of the wave function $\tilde\psi (\vec r) =
\tilde R(r)\tilde\Theta(\theta) \tilde\Phi(\varphi)$ is written in
elementary functions in the form of a standing wave,

\begin{equation}
\tilde Y_{lm}(\theta,\varphi) = \frac 1{\pi}\sqrt{\frac{2l+1}{l-|m| + 
\frac 12}}\cos\left[\left(l +\frac 12\right)\theta +
\frac\pi 2\left(l-|m|\right) \right]e^{im\varphi},   \end{equation}
where we have took into account that the phase-space integral in
the classical turning point $\theta_1$, $\chi(\theta_1)=
-\frac\pi 2(n_\theta +\frac 12)$. 

Consider now important consequences of the above solution. First of all 
note that Eq. (8) shows the existence of a nontrivial solution at $l = 0$. 
Setting in (8) $m=0$, $l=0$ we obtain

\begin{equation}
\tilde Y_{00}(\theta,\varphi) = \frac{\sqrt 2}\pi\cos\frac \theta 2.
\end{equation}
Note, that the angular eigenfunction $\tilde Y_{00}(\theta,\varphi)$ of 
the ground state is symmetric and has the form of standing half-wave. 
The corresponding eigenvalue is

\begin{equation}       M_0 = \frac{\hbar}2.     \end{equation}
The eigenvalue (10) contributes to the energy of zero oscillations.
This means that (9) and (10) can be considered as solution which
describes the quantum fluctuations of the angular momentum.

Quantization condition (7) results in the exact energy eigenvalues
for {\em all} solvable spherically symmetric potentials in quantum
mechanics. In particular, for the isotropic oscillator, we have $E_n
= \omega_0 [2\hbar(n'+\frac 12) + M]$. So far, as $M = (l+\frac
12)\hbar$, this results in the exact energy eigenvalues for the
harmonic oscillator. For the energy of zero oscillations, $E_0$, we
have $E_0 = \omega_0 (\hbar + M_0)$ that apparently shows
contribution of the quantum fluctuations of the angular momentum into
the energy of the ground state $E_0$.

The Coulomb problem is another classic example. Quantization
condition (7) again reproduces the exact result, $E_n = -\frac 12
\alpha^2m[(n_r+\frac 12)\hbar + M]^{-2}$. As in previous example, the
quantum fluctuations of the angular momentum, $M_0$, contribute to
the energy of the ground state, $E_0 = -\frac 12\alpha^2m (\frac
\hbar 2 + M_0)^{-2}$.

Third example, the Hulth\'en's potential, is of a special interest in
atomic and molecular physics. The radial problem for this potential
is usually considered at $l=0$. However, the quasiclassical approach
results in the nonzero centrifugal term at $l=0$ and allows to obtain
the analytic result for this potential at any $l$.

The leading-order WKB quantization condition (7) for the Hulth\'en
potential is

\begin{equation}
I = \int_{r_1}^{r_2}\sqrt{2m\left(E + V_0\frac{e^{-r/r_0}}
{1-e^{-r/r_0}}\right) -\frac{\vec M^2}{r^2}}dr = 
\pi\hbar\left(n^\prime +\frac 12\right).
\end{equation}
Calculation of this integral results in the energy spectrum \cite{Se}

\begin{equation}
E_n=-\frac 1{8mr_0^2}\left(\frac{2mV_0r_0^2}N-N\right)^2,
\end{equation}
where $N = (n_r+\frac 12)\hbar + M$ denotes the principal quantum
number. As in previous examples, this formula apparently shows the
contribution of the quantum fluctuations of the angular momentum to
the energy of the ground state $E_0$. Setting in (12) $M=0$, we
arrive at the energy eigenvalues obtained from known exact solution
of the Schr\"odinger equation at $l=0$. However, in our case $M_{min}
= \frac\hbar 2$ at $l=0$ and the principal quantum number is $N =
(n_r+\frac 12)\hbar + M_0$.

Thus the quasiclassical method is the appropriate method to solve the
3-dimensional Schr\"odinger's equation. The Langer correction has
a deep physical origin as the correction to the squared angular
momentum eigenvalues. It points out the existence of the integral of
motion $\vec M^2=(l+\frac 12)^2\hbar^2$. The squared angular momentum
eigenvalues $\vec M^2$ are the same for {\em all} spherically
symmetric potentials and no further corrections are necessary.

The eigenfunctions far from turning points can be written in terms of
elementary functions in the form of standing wave and can be treated
as a special class of the exact solutions. The angular eigenfunction
at $l=0$ is of the type of standing half-wave. This solution has
treated as one which describes the quantum fluctuations of the
angular momentum corresponding to the eigenvalue $M_0 =\frac\hbar 2$.


\end{document}